\newcommand{\bee}{\begin{equation}}
\newcommand{\ene}{\end{equation}}
\newcommand{\beea}{\begin{eqnarray}}
\newcommand{\enea}{\end{eqnarray}}

\documentclass[aps,preprint,pre,showpacs,superscriptaddress]{revtex4}
\usepackage{amsmath} 
\usepackage{graphicx}
\usepackage{dcolumn}
\usepackage{bm}
\begin{document}
\title{Pseudoquatum Features of Parametrically Driven Classical Fluids}
\author{M. Akbari-Moghanjoughi}
\affiliation{Faculty of Sciences, Department of Physics, Azarbaijan Shahid Madani University, 51745-406 Tabriz, Iran}

\begin{abstract}
Recent experiments on walking droplets suggest an underlying connection between fluid dynamics and the quantum mechanics. Such experiments may be used to support the de Broglie-Bohm pilot-wave theory. In this paper we show that many quantum-like features of hydrodynamic excitations can be explained in the framework of the parametrically driven nonlinear Schr\"{o}dinger equation (DNLSE). It is shown that the nonlinear Schr\"{o}dinger equation (NLSE) describes the pseudoquantum features of a given pseudoparticle in the Sagdeev pseudopotential just as the ordinary linear Schr\"{o}dinger equation (LSE) describes the quantum nature of real particles. The NLSE is shown to reduce to nonlinear Hamilton-Jacobi equation (NHJE) for the phase function of the complex wavefunction. The origin of the wave-particle dual-character in both LSE and NLSE is explained and basic similarities between the two models is highlighted. It is shown that many more pseudoquantum effects (yet to be explored) such as pseudoentanglement, self-interference effect and quantization of pseudoparticle energy are inherent features of classical fluids. Current investigation suggests that emergence of the quantum nature of fluids is independent from dynamics of the bouncing droplets or their interactions with the surface waves instantaneously created by them. In contrast to previous suggestions, it is remarked that intrinsically nonlinear collective excitations of a fluid can drive any particle even a pseudoparticle to behave quantum mechanically. These collective interactions in hydrodynamic model confirm the important nonlocal character (which is a key aspect of the de Broglie-Bohm pilot-wave theory) of all the pseudoquantum phenomena. The recent hydrodynamic quantum analog experiments and extensive new theoretical models may hopefully enable physicists to unlock the decade long hidden mystery of the quantum weirdness.
\end{abstract}

\date{\today}

\maketitle

\section{Historical Background}

It is more than a decade that walking droplets on the surface of a periodically vibrating fluid have fascinated scientists with their many pseudoquantum features \cite{cou1,cou2}. Floating of a droplet on a vibrating bath of fluid for the first time was described by Jearl Walker in 1978 in a Scientific American amateur scientist article. The recent pioneering experiments have \cite{bush1,cou3} revealed that the macroscopic millimeter-sized walking droplets on a suitably driven fluid surface can mimic variety of quantum-like effects which were already assumed to emerge at atomic and molecular scales \cite{arn}. The most prominent interference effect has been experimentally confirmed for droplets passing through single and double slit barriers \cite{pro,cou4} which are quite reminiscent to their quantum counterparts. However, at the complete contrast to the Copenhagen interpretation, the mobile droplets are observed to pass through only one slit at a time. The later effect seems to give a credit to the initially ignored deterministic pilot wave quantum theory of de Broglie \cite{de1,de2} which was then reestablished by Bohm in 1952 \cite{bohm1,bohm2}. Subsequent experiments on walking droplets have shown yet more interesting hydrodynamic quantum analog aspects such as quantum tunneling \cite{eddi}, bound state orbital quantization \cite{fort1,oza,harris} and Landau levels \cite{fort2}. Therefore, classical experiments with tiny magic droplets seem to fundamentally alter our conceptions of wave-particle duality and consequently foundations of quantum mechanics itself in near future \cite{zee}.

On the other hand, while the novel classical analogy between quantum particles and bouncing droplets may have shed light on the understanding of quantum weirdness, the underlying physical and mathematical connections between de Broglie-Bohm pilot-wave and the hydrodynamic theories is still a matter of debate \cite{chris}. It is however apparent that the akin similarities between quantum wave-particle phenomenon and the interacting droplet with its own hydrodynamic wave is pointing at a much deeper physical description of the quantum mechanics. The later suggestion has motivated an intensified research \cite{gil,hu,harris2,oza1,oza2,oza3,bush2,mil,bush3,lab,puc,burn,tam} towards detailed theoretical and experimental investigation of the dynamics of magic bouncing droplet and its interaction with the surface waves produced by the vibrating platform. Unfortunately, not much of a progress has been made towards a unified theoretical description of the phenomenon due to the very complex nature of collective interactions in fluids and the rapid temporal evolution of the droplet specific parameters. In current research we show that in order to investigate the hydrodynamic quantum analog problem there is no need to study the dynamics of the bouncing droplet since as will be shown it is the fluid (hands of a magician) itself that does the magic to whatever object (which we consider a ghostly one named the pseudoparticle) that floats on the surface perturbations. Present theory also reveals that nonlinear hydrodynamic excitations are capable of producing many other fundamental pseudoquantum features without considering even a real particle dynamics.

\section{Dynamics of Hidden Particles}

Fluids including plasmas accommodate important nonlinear excitations such as solitons and cnoidal waves which carry energy in fundamentally different manner than ordinary linear perturbations. One of the distinguished characteristics of nonlinear excitations is their unique dependence of oscillation amplitude to frequency which is not observed for sinusoidal waves. The later describes many interesting features of the nonlinear electrostatic waves in an electron-ion plasma such as energy spectrum, nonlinear resonance, harmonic generation, chaotic excitations and autoresonance effects which has been recently revisited in Refs. \cite{akbarienergy,akbariresonance,akbariharmonic,akbariext,akbarishock,akbariauto} using the recently developed pseudoparticle dynamics method. In the stationary frame $\xi=x-Mt$ ($M$ being the Mach number), the dissipationless hydrodynamic equations can be reduced to the following pseudoforce (Newton) equation of motion \cite{akbariforce}
\begin{equation}\label{pseudoforce}
\frac{{{\partial ^2}\Phi }}{{\partial {\xi ^2}}} - F(\Phi ) = 0,
\end{equation}
\begin{figure}[ptb]\label{Figure1}
\includegraphics[scale=.48]{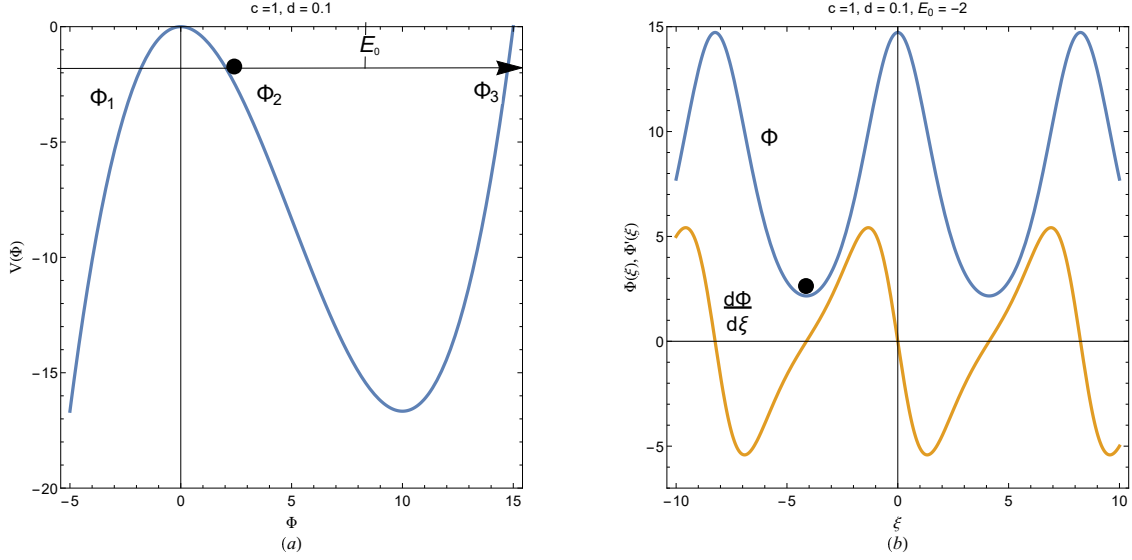}\caption{(a) The Helmholtz ($p=1$) pseudopotential for given values of $c$ and $d$ with the pseudoparticle shown at one of turning points (b) The corresponding pseudoparticle pseudotime evolution and its pseudospeed for the given pseudoenergy value $E_0=-2$.}
\end{figure}
in which $\Phi$ may describe the electrostatic field for plasmas, gravitational field for surface waves or other appropriate force fields. Also, $F(\Phi)$ is a conservative force corresponding to the Sagdeev pseudopotential $V(\Phi ) =  - \int {F(\Phi )} d\Phi$ in which the unit-mass pseudoparticle oscillates. In this formulation $\Phi$ and $\xi$ play the analogous roles of position $x$ and time $t$, respectively. Equation (\ref{pseudoforce}) can be written in the form of energy equation as follows
\begin{equation}\label{energy}
\frac{1}{2}\left( {\frac{{\partial \Phi }}{{\partial \xi }}} \right)^2 + V(\Phi ) = {E_0},
\end{equation}
in which $E_0$ is the conserved total energy of the pseudoparticle. The pseudoparticle formalism is particularly interesting because various collective excitations in a fluid even in the presence of dissipation \cite{akbarishock} and external forcing effect can be described in terms of dynamics of a single particle-like motion in the generalized Sagdeev pseudopotential. In the small-amplitude weakly nonlinear potential limit of form $V(\phi)=-(c/2)\Phi^2+(d/3)\Phi^3$ or $V(\phi)=-(c/2)\Phi^2+(d/3)\Phi^4$ which are known as the Helmholtz and Duffing potentials, respectively, the energy equation (\ref{energy}) can be fully integrated to give $\Phi(\xi)$ in terms of Jacobi-elliptic functions \cite{akbarienergy}. The small-amplitude weakly nonlinear waves in a fluid are described by the celebrated Korteweg de Vries equation (KdVE) given as
\begin{equation}\label{diff}
\frac{{\partial \Phi(x,t)}}{{\partial t}} + \frac{{{\partial ^3}\Phi(x,t)}}{{\partial {x^3}}} + d\Phi(x,t)\frac{{\partial \Phi^p(x,t)}}{{\partial x}} = 0,
\end{equation}
with nonlinear coefficient $d$ which for a traveling waves with speed $c$ ($\xi=x-ct$) for $p=\{1,2\}$ reduces to the energy equation of the form (\ref{energy}) with the Helmholtz and Duffing potentials, respectively. The small-amplitude cnoidal wave solution for the case of $p=1$ is given as
\begin{equation}\label{cnoidal}
\Phi (\xi ) = {\Phi _2} + \left( {{\Phi _3} - {\Phi _2}} \right){\rm{c}}{{\rm{n}}^2}\left[ {\sqrt {\frac{{{{d}}\left( {{\Phi _3} - {\Phi _1}} \right)}}{{6}}} \xi ,\frac{{\left( {{\Phi _3} - {\Phi _2}} \right)}}{{\left( {{\Phi _3} - {\Phi _1}} \right)}}} \right],
\end{equation}
where $\Phi_1<\Phi_2<\Phi_3$ are three solutions of $E_0-V(\Phi)=0$ given by
\begin{equation}\label{roots}
{\Phi _{1,2}} = \frac{c}{{2d}} - \frac{{\left( {1 \pm {\rm{i}}\sqrt 3 } \right){c^2}}}{{4dT}} - \frac{{\left( {1 \mp {\rm{i}}\sqrt 3 } \right)T}}{{4d}},\hspace{3mm}{\Phi _3} = \frac{1}{2}\left( {\frac{c}{d} + \frac{{{c^2}}}{{dT}} + \frac{T}{d}} \right),
\end{equation}
where the coefficient $T$ given as below
\begin{equation}\label{t}
T = {\left[ {{c^3} + 12{d^2}E_0 + 2\sqrt 6 \sqrt {{c^3}{d^2}E_0 + 6{d^4}E_0^2} } \right]^{1/3}}.
\end{equation}
Figure 1(a) shows the Helmholtz pseudopotentialfor given values of $c=1$ and $d=0.1$. The total energy level $E_0=-2$ is shown by horizontal arrow and the pseudoparticle is shown by a filled circle at one of the turning points. The pseudoparticle can oscillate in the range $\Phi_2<\Phi<\Phi_3$ in this configuration. Figure 1(b) shows the pseudotime evolution of the corresponding pseudoparticle and its pseudospeed. It is remarked that the motion of the epseudoparticle is truly nonlinear and its speed maximized at the bottom of the pseudopotential well and becomes zero at turning points. Note that in stationary frame $\xi$ the pseudoparticle can be assumed to sweep the wave crust with a horizontal constatnt speed $c$ and moves up and down with a variable speed $d\Phi/d\xi$ shown in Fig. 1(b). The pseudoparticle shown in Fig. 1 is completely imaginary and has nothing to do with the bouncing droplet of the above mentioned hydrodynamic quantum analog experiments. However, it is quite realistic to imagine that the falling droplet interacts dynamically with this pseudoparticle with well-defined position and speed $v=\sqrt{c^2+(d\Phi/d\xi)^2}$. It is remarked that hydrodynamic theory may be considered as nonlocal just like the pilot wave theory by assuming the parameters of the pseudoparticle to be the hidden variables of this fluid model.

\section{LSE Versus NLSE}

The Linear Schr\"{o}dinger Equation (LSE) describes the time evolution of a physical system in which quantum effects like wave-particle duality play the dominant role. It is a Hamilton-Jacobi equation (HJE) which describes the energy levels of real quantum particles. Using the ikonal wavefunction $\Psi(x,t)=R(x,t)\exp[iS(x,t)]$ in the LSE
\begin{equation}\label{lse}
\left[ {i\hbar \frac{\partial }{{\partial t}} + \frac{{{\hbar ^2}}}{{2m}}\left(\frac{{{\partial ^2}}}{{\partial {x^2}}}\right) - U(x)} \right]\Psi (x,t) = 0,
\end{equation}
leads to the following HJE
\begin{equation}\label{hje}
\frac{{\partial S}}{{\partial t}} + \frac{1}{{2m}}{\left( {\frac{{\partial S}}{{\partial x}}} \right)^2} + {U_g} = 0,\hspace{3mm}{U_g} = U(x) - \frac{{{\hbar ^2}}}{{2m}}\left( {\frac{1}{R}\frac{{{\partial ^2}R}}{{\partial {x^2}}}} \right),
\end{equation}
in which $U_g$ is a generalized potential including the Bohm's quantum potential. In this representation of wavefunction the functions $R(x,t)$ and $S(x,t)$ behave quite differently. It is remarked that the phase function $S(x,t)$ behaves little like waves \cite{hje} but $R(x,t)$ does not. Madelung \cite{mad} has used this representation for the first time in order to derive the so-called quantum hydrodynamics including an additional quantum potential. The later addition has initiated a new field known as the deterministic pilot-wave version of quantum mechanics \cite{dur1,dur2,dur3}. The same procedure as Madelung has been employed to derive the quantum hydrodynamic and kinetic formulations for plasmas \cite{haasbook}.

The nonlinear Schr\"{o}dinger Equation (NLSE), on the other hand, is used to model slow modulation of wave trains like cnoidal perturbation in weakly nonlinear fluid-like environments. More recently, the traveling-waves solution of the NLSE with arbitrary nonlinearity has been derived in Ref. \cite{akbarinlse}. Such a generalized NLSE reads
\begin{equation}\label{nlse}
\left\{ {i\frac{\partial }{{\partial t}} + \frac{{{\partial ^2}}}{{\partial {x^2}}} + f\left[ {\left| {\Psi (x,t)} \right|} \right]} \right\}\Psi (x,t) = 0,
\end{equation}
in which $\Psi (x,t)$ is the complex wavefunction of real variables and $f\left[ {\left| {\Psi (x,t)} \right|} \right]$ is the real generalized nonlinear function. In Ref. \cite{akbarinlse} it has also been shown that (\ref{nlse}) and the generalized KdVE share the same energy spectrum for the corresponding pseudoparticles. While the NLSE can not describe real particle dynamics in real potential it can be used to evaluate the dynamics of a pseudoparticle trapped in the Sagdeev pseudopotential, as follows. It is noted that the NLSE like LSE can be converted into HJE for the phase function $S(x,t)$ in the above-mentioned Madelung representation as below
\begin{equation}\label{hje}
\frac{{\partial S}}{{\partial t}} + {\left( {\frac{{\partial S}}{{\partial x}}} \right)^2} - {f_g} = 0,\hspace{3mm}{f_g} = f(R) + \frac{1}{R}\frac{{{\partial ^2}R}}{{\partial {x^2}}},
\end{equation}
which may be referred to as the nonlinear Hamilton-Jacobi equation (NHJE). Note that NHJE is related to HJE through definitions $\hbar=1$, $m=1/2$ and $U(x)\to -f(R)$. Note also that, the NLSE in the reduced stationary wave coordinate ($\xi=x-ct$) can be converted to the more fundamental pseudoforce equation given by Eq. (\ref{pseudoforce}) as follows. By setting a solution $\Psi(R,\xi)=R(\xi)\exp[ic\xi/2]$ in NLSE (\ref{nlse}) we find the following pseudoforce equation
\begin{equation}\label{npf}
\frac{{{\partial ^2}\Phi }}{{\partial {\xi ^2}}} - F(\Phi ) = 0,\hspace{3mm}F(\Phi ) = -\left[ {\frac{{{c^2}}}{4} + f(\Phi )} \right]\Phi,
\end{equation}
in which $F(\Phi)$ is the corresponding pseudoforce. Therefore, the modulus of the wavefunction describes the dynamics of a pseudoparticle oscillation in the Sagdeev potential $V(\Phi ) =  - \int {F(\Phi )} d\Phi$. The complex wavefunction for the Helmholtz oscillator $p=1$ reads
\begin{equation}\label{helmsolr}
\Psi (\Phi,\xi) = \left\{ {{\Phi _2} + \left( {{\Phi _3} - {\Phi _2}} \right){\rm{c}}{{\rm{n}}^2}\left[ {\sqrt {\frac{{d\left( {{\Phi _3} - {\Phi _1}} \right)}}{6}} \xi,\frac{{\left( {{\Phi _3} - {\Phi _2}} \right)}}{{\left( {{\Phi _3} - {\Phi _1}} \right)}}} \right]} \right\}\exp \left( {\frac{{\pm ic\xi}}{2}} \right),
\end{equation}
with $\xi=x\pm ct$ and plus/minus sign denoting the right/left traveling nonlinear wave. The imaginary part of the stationary wavefunction represents a linear wave which describes the wave-like behavior of the pseudoparticle just like real particles in LSE. This feature will be elucidated in the next section where the wave-particle duality of NLSE is considered. Note that in the stationary frame ($\xi=x\pm Mt$) (\ref{nlse}) can fully describe the pseudoquantum features of the pseudoparticle trapped in the given Sagdeev potential $V(\Phi ) = \int {f(\Phi )\Phi d\Phi  + {M^2}{\Phi ^2}/8}$ which can be directly obtained from any dissipationless hydrodynamic set of equations \cite{akbariforce}. Therefore, the pseudoquantum features of hydrodynamic equations can be studied in a stationary frame using the following generalized pseudo-NLSE (pNLSE) \cite{akbarinlse}
\begin{equation}\label{pnlse}
\left[ {\pm iM\frac{\partial }{{\partial \xi }} + \frac{{{\partial ^2}}}{{\partial {\xi ^2}}} + f(\Phi )} \right]\Psi (\Phi,\xi) = 0.
\end{equation}
Note that Eq. (\ref{pnlse}) has similar relation to the pseudoforce equation (\ref{pseudoforce}) just as the Schr\"{o}dinger equation does to the Newton's second law of motion. The pNLSE is also related to the pseudo-HJE (pHJE) through the following relations
\begin{equation}\label{phje}
\pm M\frac{{\partial \Lambda}}{{\partial \xi}} + {\left( {\frac{{\partial \Lambda}}{{\partial \xi}}} \right)^2} - {W_g} = 0,\hspace{3mm}{W_g} = f(\Phi) + \frac{1}{\Phi}\frac{{{\partial ^2}\Phi}}{{\partial {\xi^2}}},
\end{equation}
via the Madelung transformation $\Psi(\Phi,\xi)=\Phi(\xi)\exp[i\Lambda(\xi)]$. Note that the real part of the wave function satisfies the pseudoforce equation (\ref{npf}). It is observed that in (\ref{pnlse}) $f(\Phi)$ plays the analogous role of the $x$-dependent potential $U(x)$ in the LSE (\ref{lse}) and in (\ref{phje}) $W_g$ plays the generalized potential including the pseudoquantum potential $Q=(1/\Phi)(\partial^2 \Phi/\partial \xi^2)$. We may also write the analogous parametrically driven pseudo-NLSE (pDNLSE) as \cite{bar1}
\begin{equation}\label{pdnlse}
\left[ {\pm iM\frac{\partial }{{\partial \xi }} + \frac{{{\partial ^2}}}{{\partial {\xi ^2}}} + f(\Phi )} \right]\Psi (\Phi,\xi ) - \mathcal{F}(\xi ){\Psi ^*}(\Phi,\xi ) = 0,
\end{equation}
in which $\Psi^*$ is the complex conjugate of $\Psi$ and $\mathcal{F}(\xi)$ is the generalized driving force. Equation (\ref{pdnlse}) will be referred to in next section. The externally driven pseudo-NLSE is given as \cite{bar1}
\begin{equation}\label{xpdnlse}
\left[ {\pm iM\frac{\partial }{{\partial \xi }} + \frac{{{\partial ^2}}}{{\partial {\xi ^2}}} + f(\Phi )} \right]\Psi (\Phi,\xi ) - \mathcal{F}(\xi )= 0.
\end{equation}
Finally, we can write the damped externally-driven pseudo-NLSE (pDDNLSE) as \cite{bar1}
\begin{equation}\label{xpddnlse}
\left[ {\pm iM\frac{\partial }{{\partial \xi }} + \frac{{{\partial ^2}}}{{\partial {\xi ^2}}} + f(\Phi ) + i\Gamma } \right]\Psi (\Phi,\xi ) - \mathcal{F}(\xi ) = 0,
\end{equation}
in which $\Gamma>0$ denotes the damping coefficient. In (\ref{pdnlse}), (\ref{xpdnlse}) and (\ref{xpddnlse}) the term $\mathcal{F}(\xi)$ denotes a generalized force which for periodic drive it can be written as $\mathcal{F}_0\exp(il\Omega\xi)$ in which $\mathcal{F}_0$ and $\Omega$ denote the external driving force magnitude and frequency, respectively. Also, the parameter values of $l=\{1,2\}$ is used for external and parametric driving forces, respectively. Note that in the reduced $\xi=x\pm Mt$ system the roles played by the time $t$ and the space $x$ coordinates in NLSE has transferred to the $i\xi$ and $\xi$, respectively.

\section{Wave-Particle Duality}

\begin{figure}[ptb]\label{Figure2}
\includegraphics[scale=.55]{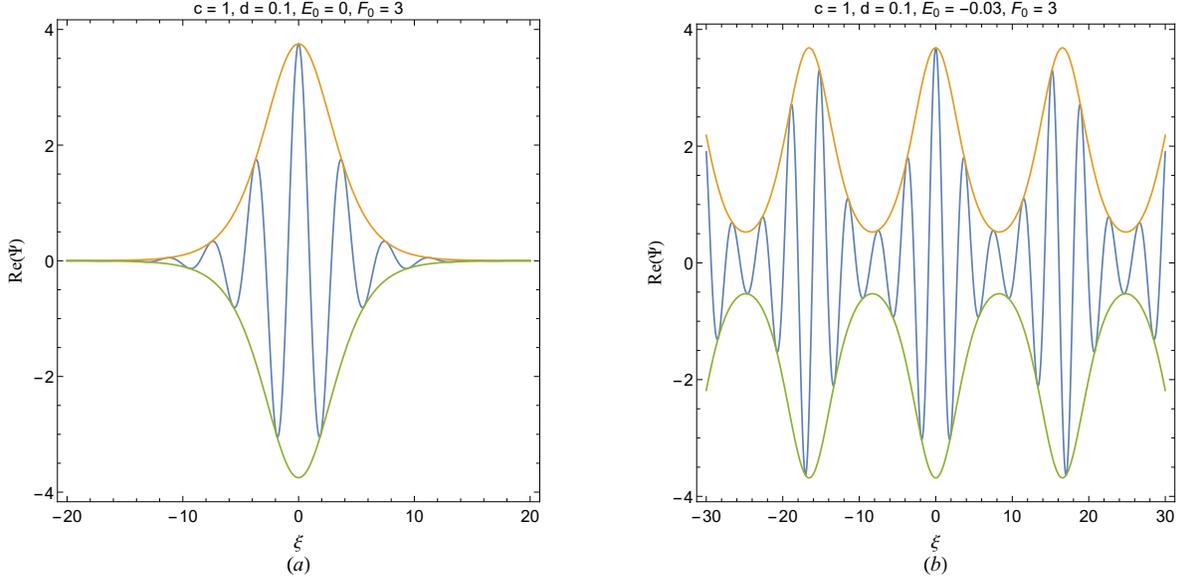}\caption{(a) The traveling wave solution of DNLSE for envelope solitonic excitation with $E_0=0$. (b) The travaling wave solution of DNLSE for envelope cnoidal wave excitation with $E_0=-0.03$ and similar other parameters. The envelope curve is given by the Jacobi-elliptic function in (\ref{helmsolr}).}
\end{figure}
One of the fundamental aspects of the quantum weirdness arises from wave-particle duality. But how the LSE can describe both the wave and particle aspects of the quantum entities? How can the walking droplets mimic such a novel property in a macroscopic double-slit experimental setup? These are some of the main questions to be answered in this section. In the previous section we clarified some basic similarities between LSE and NLSE. It turns out however that these two fundamental equations have more in common than their apparent formulations. To this end, let us consider a complex differential equation $\mathcal{L}\Psi(x,t)=0$ with $\mathcal{L}$ being the complex operator and $\Psi(x,t)$ a complex function. For the LSE operator in (\ref{lse}) we have $\mathcal{L}=i\hbar\partial/\partial t+(\hbar^2/2m)\partial^2/\partial x^2-U(x)$. Such a differential equation has complex solutions like $\Psi(x,t)=u(x,t)+iv(x,t)$. It is noted that such a solution has a unique property of different behavior along imaginary and real coordinates, simultaneously, since both functions $u(x,t)$ and $v(x,t)$ should satisfy the complex differential equation. This is called a double solution differential equation. And the de Broglie-Bohm pilot-wave double solution theory originates from this unique property of the LSE. Note that the complex solution may be written in the polar form $\Psi(x,t)=R(x,t)\exp[iS(x,t)]$ with $R(x,t)=\sqrt{u(x,t)^2+v(x,t)^2}$ and $S(x,t)=\arctan[v(x,t)/u(x,t)]$ which is the Madelung representation of a wavefunction. For instance for the problem of a particle in a box the LSE gives rise to a combined solution of $\Psi(x,t)=r(x)s(t)$ in which $r(x)$ is a real function describing the particle-like nature (oscillation of the particle trapped in the one-dimensional box) along the spacial coordinate and $s(t)=\exp(iEt/\hbar)$ is complex function describing the wave behavior of the particle with the quantized energy E along the time coordinate. It is remarkable that the given quantum entity described by such equation must have dual character with particle-like motion in spacial dimension and wave-like behavior along the time coordinate. It is evident that the NLSE (\ref{nlse}) solution has also dual character as is reflected in the solution (\ref{helmsolr}). In order to be more clear let us consider the parametrically driven NLSE (DNLSE) \cite{bar1,bar2,bar3,bar4} with a periodic external bias as follows
\begin{equation}\label{dnlse}
\left\{ {i\frac{\partial }{{\partial t}} + \frac{{{\partial ^2}}}{{\partial {x^2}}} + f\left[ {\left| {\Psi (x,t)} \right|} \right]} \right\}\Psi (x,t) - \mathcal{F}(x,t){\Psi ^*}(x,t) = 0,
\end{equation}
in which $\Psi^*$ is the complex conjugate of $\Psi$ and $\mathcal{F}(x,t)$ is the generalized driving amplitude. The DNLSE (\ref{dnlse}) describes the slow modulation of nonlinear excitations in a fluid-like medium. With a periodic driving force $\mathcal{F}(x,t)$ it also describes the well-known walking droplet experiment situation. The traveling wave solution of pDNLSE (\ref{pdnlse}) has been recently investigated in Ref. \cite{akbaridnlse}. Here we consider the modulation at subcritical speed for the quadratic potential nonlinearity. Therefore, we find the following solution
\begin{equation}\label{dnlsesub}
\Psi (\Phi,\xi ) = \left\{ {{\Phi _2} + \left( {{\Phi _3} - {\Phi _2}} \right){\rm{c}}{{\rm{n}}^2}\left[ {\sqrt {\frac{{d\left( {{\Phi _3} - {\Phi _1}} \right)}}{6}} \xi ,\frac{{\left( {{\Phi _3} - {\Phi _2}} \right)}}{{\left( {{\Phi _3} - {\Phi _1}} \right)}}} \right]} \right\}\exp \left( {i\xi \sqrt {{F_0} - \frac{{{c^2}}}{4}} } \right),
\end{equation}
in which $F_0$ is the driving force magnitude and the pseudopotential roots are given as
\begin{equation}\label{rootshelm1}
{\Phi _1} = \frac{1}{8}\left( {\frac{{{c^2}}}{d} - \frac{{{c^4}}}{{dT}} + \frac{T}{d}} \right),\hspace{3mm}{\Phi _{2,3}} = \frac{{{c^2}}}{{8d}} + \frac{{\left( {1 \mp {i}\sqrt 3 } \right){c^4}}}{{16dT}} + \frac{{\left( {1 \pm {i}\sqrt 3 } \right)T}}{{16d}},
\end{equation}
where
\begin{equation}\label{t1}
T = {\left( {16\sqrt 6 \sqrt {{c^6}{d^2}{E_0} + 384{d^4}{E_0}^2}  - 768{d^2}{E_0} - {c^6}} \right)^{1/3}}.
\end{equation}
The solution (\ref{dnlsesub}) has the same structure as (\ref{helmsolr}) with the difference that in the former $F_0$ provides more control over the carrier wave frequency. The fundamental difference between the solution (\ref{dnlsesub}) of the reduced ($\xi=x\pm ct$) DNLSE with that of the LSE $\Psi(x,t)=r(x)s(t)$ is that in the former both particle and wave-like feature of the solution can be observed along the reduced coordinate while for the later the simultaneous detection of both features is impossible due to the principle of complementarity. In Fig. 2 we show the traveling wave solution of the DNLSE. Figure 2(a) shows the envelope soliton excitation given by (\ref{dnlsesub}) for $E_0=0$. The envelope curve indicates the nonlinear fluid excitations on which the imaginary pseudoparticle slides up and down. This curve shows the particle-like character of the fluid excitation. Zabusky and Kruskal \cite{zk} for the first time discovered a remarkable particle like character of solitons by numerical simulations when nobody appreciated the prominent importance of nonlinearity. They showed that solitons as nonlinear wave solution of KdV equation unlike their linear counterparts can ghostly collide and pass through each other without a significant trace in their original shapes other than a slight phase shift. Figure 2(b) shows envelope cnoidal excitation with the same carrier wave frequency as in Fig. 2(a).

\section{Pseudoentaglement}

\begin{figure}[ptb]\label{Figure3}
\includegraphics[scale=.55]{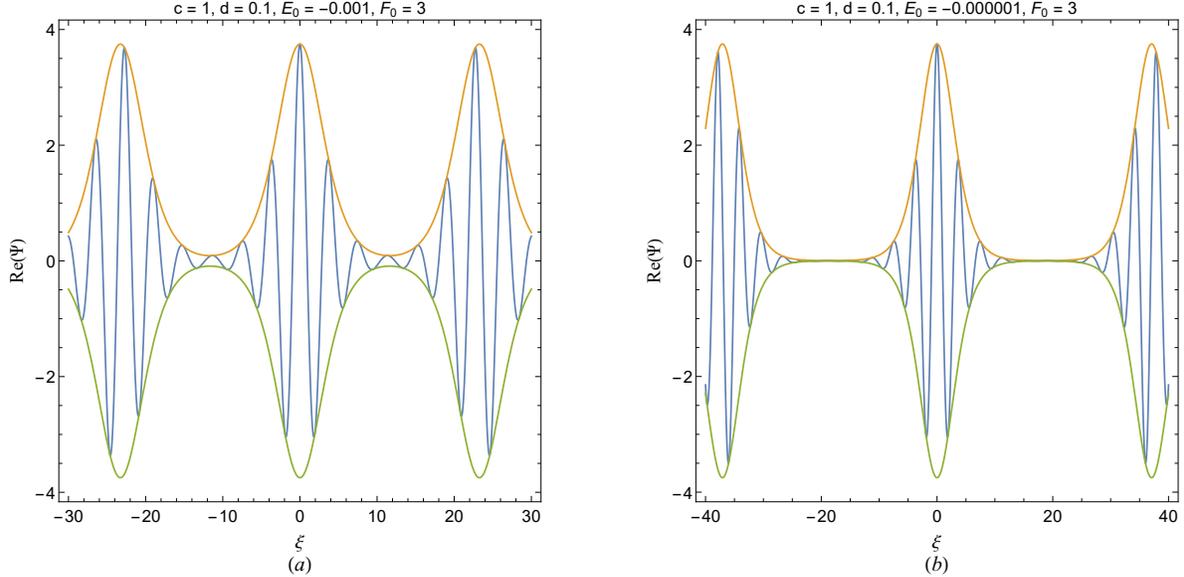}\caption{(a) The entangled envelope solitons with energies $E_0=-0.001$ (b) The entangled envelope solitons with energies $E_0=-0.000001$.}
\end{figure}

Quantum entanglement is one of many peculiar predictions of the quantum theory for correlated particles. It occurs when pair or group of particles interact in such a way that the alteration of quantum state of one changes the state of others in the system. This is one of the weirdest properties of a quantum system which is usually referred to as an action from a distance and became a matter of intense debate after the 1935 paper by Albert Einstein, Boris Podolsky, and Nathan Rosen \cite{epr}, what has became known as the EPR paradox since then. Einstein and some other opponents of the quantum theory considered this spooky action at a distance in direct violation of local realist view of casualty and assumed this prediction as a sign for incomplete quantum formulation. Their reasoning also included the superluminal communications which was in direct contrast to the special theory of relativity. The later has caused dissatisfaction of Schr\"{o}dinger himself of the phenomenon. Consequent experiments on polarization and spin of particles however confirmed this prediction of the quantum theory. This and many other quantum weird phenomena seemed so unnatural at the time that the well-known physicist Richard Feyneman puts it in some words \cite{nat}: “Nature isn't classical, dammit, and if you want to make a simulation of nature, you'd better make it quantum mechanical, and by golly it's a wonderful problem, because it doesn't look so easy.” But obviously many recent classical experiments on the walking droplets are all proofs against this famous quotation. Now going back to the solution of DNLSE, it is found that envelop humps in Fig. 2(b) which act like particles can be made arbitrarily far from each other by making the pseudoparticle energy $E_0<0$ very close to zero. In doing so we are able to make envelope solitons which are all solutions of the same DNLSE equation and hence entangled. Figure 3 shows the DNLSE solution for values of the energy as $E_0=-0.001$ and $E_0=-0.000001$ with same other parameter values in plots 3(a) and 3(b), respectively. It is clearly remarked that the distance between envelope solitons can be made arbitrarily large by appropriate setting of the value of $E_0$. The later confirms that the entanglement phenomenon is classical in nature and can happen in macroscopic systems. There is yet another interesting fact about the entanglement and that is it can happen in two distinct forms, namely, spacial and temporal entanglements. The spacial entanglement is related to the well-known Akhmediev breather type solutions \cite{akh} and the temporal entanglement to the Kuznetsov-Ma breather type solutions \cite{km} of the DNLSE.

\section{Interference Effect}

\begin{figure}[ptb]\label{Figure4}
\includegraphics[scale=.55]{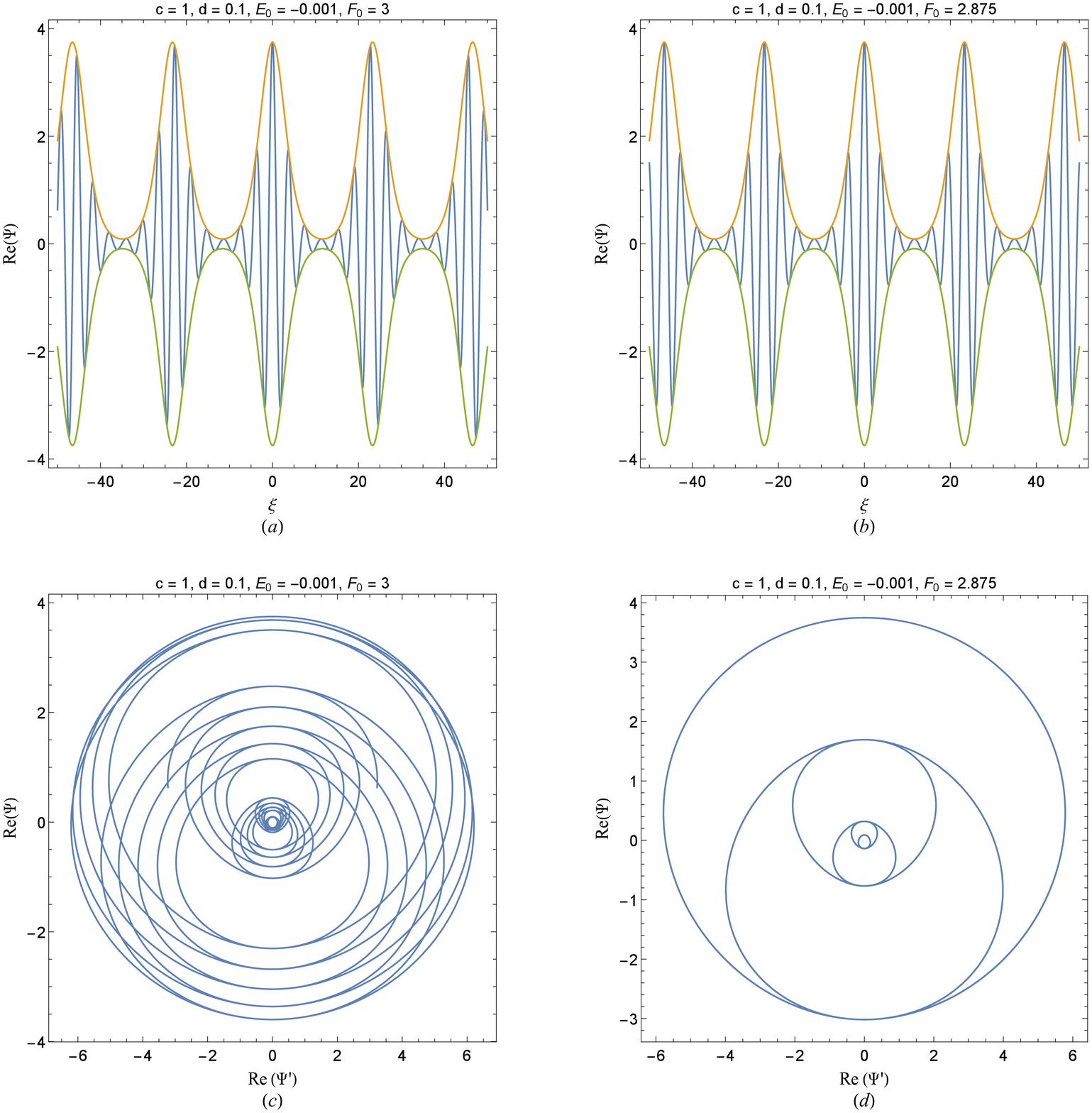}\caption{(a) and (b) The envelope cnoidal excitation with same given parameters and two different driving force manitude. (c) and (d) The corresponding phase space trajectories for plots (a) and (b). The resonance occurs for the driving force value of $F_0=2.875$.}
\end{figure}
The great physicist Richard Feynman once quoted that \cite{rf} "the double-slit experiment has in it the heart of quantum mechanics. In reality, it contains the only mystery". The phenomenon has become a representative icon of quantum mechanics theory, since its birth. Recent experiments with walking droplets however proved that \cite{pro,cou3} the phenomenon is not exclusive feature of the microscopic quantum systems. However, these experiments may be used to support \cite{bush3} the alternative version of quantum mechanics, namely, the de Broglie Bohm pilot wave theory. The ambiguity in the later version of the quantum mechanics however arises from the fact that which wave is guiding the particle. In the walking droplet interference experiments the droplet is assumed to be piloted by the wave which is created due to interaction of the droplet with the liquid surface. No such an interaction can be traced in the quantum theory of particles. Therefore, one is indulged to work with a ghostly de Bloglie wave which is attached to and moves with every quantum particle.

According to the hydrodynamic theory of oscillations the pseudoquantum interference effects can occur in two different ways. The first is the self-interaction of the driven nonlinear excitations. It is expected that self-interference effect to occur when the pseudofrequency of the nonlinear envelope matches that of the carrier wave. This is to say
\begin{equation}\label{self}
\sqrt {\frac{{d\left( {{\Phi _3} - {\Phi _1}} \right)}}{6}}  = n\sqrt {{F_0} - \frac{{{c^2}}}{4}},
\end{equation}
in which $n$ is an integer. For instance for given values of $c$, $d$ and $E_0$ parametric variation of the driving force strength $F_0$ can setup the self interference effect for appropriate values. Such an interference may be visualize as the harmony between the motion of the pseudoparticle in the Sagdeev pseudopotential and the driving force. This kind of interference is shown in Fig. 4. It is observed that for value of $F_0=3$ the trajectory in phase space is not closed. However, as this value decreases to $F_0=2.875$ the self-interference sets in by closing the trajectory shown in Fig. 4(d). The self-interference effect leads the effect called the autoresonance-like effect \cite{akbariauto} in which the external drive leads to amplification of the nonlinear excitations.

The second type of the interference is the double slit setup. Considering a fluid in a two-dimensional container with a double slit barrier in the middle. Driving periodically fluid in one side of the wall will induce the double slit interference effect in the other side. It is evident that each slit plays a driving source with its opening size being related to the forcing amplitude. In a point of distances $\xi_1$ and $\xi_2$ from slits will be constructive if the wavelength of carrier waves match the relative distance of the point from sources, i.e., if $|\xi_2-\xi_1|=n\lambda$ in which $n$ is an integer and $\lambda$ is the wavelength of the envelope excitations. Note that in current theory the double slit interference effect of DNLSE solutions does not require considering the dynamics of any real particle. The walking droplets on the surface of fluid in experiments are guided by the interference (electrostatic or other field) patterns which are instantly formed on the wall by the driving sources before a walking droplet arrives. We speculate here that this is also the situation for the real quantum interference effect. If the later assertion is true, it will then alter our most basic conception of the quantum theory. Quite different from both orthodox and deterministic theories of the quantum mechanics, in current hydrodynamic model individual particles are guided by their own collective excitations. The later phenomenon is quite analogous to the well-known Landau damping phenomenon in which the fluid particles interact with hydrodynamic waves. The hydrodynamic model is also leads to the empty wave characteristics noted in pilot wave theory in which the true wavefunction is not influenced by individual particles \cite{bell}.

\section{Pseudoenergy Quantization}

\begin{figure}[ptb]\label{Figure5}
\includegraphics[scale=.55]{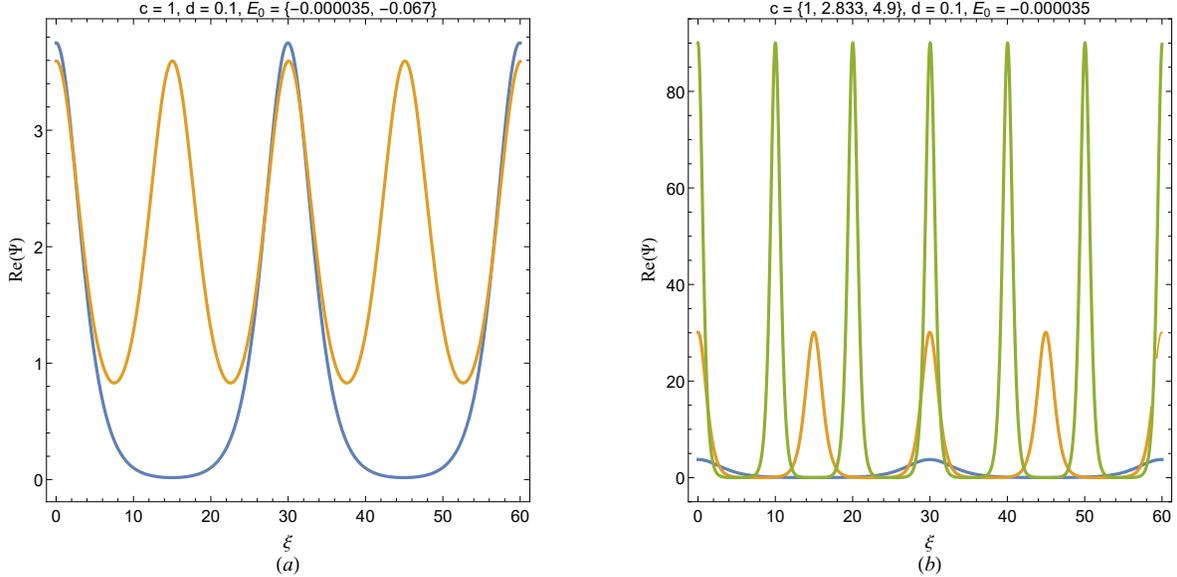}\caption{(a) Quantization of pseudoenergy of the pseudoparticle for given values of $c$ and $d$. (b) Degeneracy of the wavefunction for the given pseudoenergy $E_0=-0.000035$, different values of $c$ and same other values.}
\end{figure}
We now turn into the problem of the fluid in a one-dimensional box analogous to the famous problem of particle in a 1D box with quantized energy levels. It is evident the DNLSE (\ref{dnlse}) with a periodic forcing $\mathcal{F}(t)=F_0\exp(2i\Omega t)$ admits the standing wave solution of the form $\Psi(x,t)=\Phi(x)\Lambda(t)$ in which $\Phi(x)$ and $\Lambda(t)$ are given by
\begin{equation}\label{standing}
\Phi (x) = {\Phi _2} + \left( {{\Phi _3} - {\Phi _2}} \right){\rm{c}}{{\rm{n}}^2}\left[ {\sqrt {\frac{{d\left( {{\Phi _3} - {\Phi _1}} \right)}}{6}} x,\frac{{\left( {{\Phi _3} - {\Phi _2}} \right)}}{{\left( {{\Phi _3} - {\Phi _1}} \right)}}} \right],\hspace{3mm}\Lambda (t) = \exp(i\Omega t),
\end{equation}
in which the roots $\Phi_i$ are given by (\ref{roots}) with $c=F_0+\Omega$. Such parametric driven nonlinear standing waves are called Faraday excitations \cite{wang1,wang2,wang3,wang4}. If we consider the case of an electron-ion plasma with a conducting box then the electrostatic field vanishes at the walls of the container which are assumed to be at $x=0$ and $x=L$. Then the pseudoenergy $E_0$ of the pseudoparticle motion in the Sagdeev pseudopotential becomes quantized as
\begin{equation}\label{quant}
n\lambda ({E_0}) = 2n\sqrt {\frac{6}{{d\left( {{\Phi _3} - {\Phi _2}} \right)}}} K\left[ {\frac{{\left( {{\Phi _3} - {\Phi _2}} \right)}}{{\left( {{\Phi _3} - {\Phi _1}} \right)}}} \right] = 2L,
\end{equation}
in which $\lambda$ denotes the wavelength of the nonlinear excitations, $n$ is an integer and $K(m)$ is the elliptic integral of first kind. Note that for nonlinear hydrodynamic excitations the wavelength $\lambda(E_0)$ is interrelated to the oscillation amplitude $A(E_0)=\Phi_3-\Phi_2$. Therefore, for Faraday waves the amplitude of oscillations is also quantized. This double quantization of fluid in a box produces the spectacular Faraday patterns in a two dimensional surface. Note that the energy levels of the pseudoparticle in the limit of $E_0\simeq V_{min}$ ($V_{min}$ being the minimum of the Sadeev potential) reduces to that of a harmonic oscillator with $\hbar=1$ and $m=1/2$. Figure 5(a) shows the quantization of the pseudoenergy $E_0$ for given values other parameters. On the other hand, Fig. 5(b) shows different wavefunctions with different values of parameter $c$ can coexist with the same energy values which is equivalent to degeneracy. The nonlinear waves in Fig. 5 look quite similar to the periodic faraday waves of Ref. \cite{bush3}.

\section{Quantum Fluids}

The standard quantum theory despite its overwhelming success in many areas of science, has encountered fundamental misinterpretations since its discovery. The first objection to Copenhagen-style probabilistic interpretation of quantum measurements proposed by Max Born and others raised by Einstein quoting "God does not play dice with the universe". The deterministic pilot wave theory of de Broglie was abandoned due to several objections in the Solvay conference. However, by passing time a need for a real world interpretation of the quantum theory became necessary. In 1925 Bohm independently developed his hidden variable pilot wave theory by formulating a new dynamical law which governs the quantum particles. The law of motion given by Bohm is explicitly nonlocal meaning that the position of each particle in the system depends on every other particle in the universe. In the pilot wave theory the exact origin of the guiding wave is not clear. The strong belief on the Schr\"{o}dinger equation has driven many physicists towards different philosophical interpretations of quantum mechanics and LSE, notably the Many World Interpretation (MWI). In Spontaneous Collapse Theory (SCT), namely the one first proposed by Gian Carlo Ghirardi, Alberto Rimini, and Tullio Weber (GRW) in 1986, it is assumed that the LSE can not give a complete description of a quantum system due to its linear structure.

The origin of confusion regarding the physical interpretation of quantum mechanics may be related to the linear structure of LSE and its applications for the case of a single particle instead of ensemble of particles. Such a confusion, for instance, arises when one realizes a single particle double-slit experiment, expecting a wave-like pattern formation or expects the quantum tunneling to occur for a single electron which is impossible to understand classically. These are in fact some of misinterpretations of LSE regarding probabilistic results for single particles. It should however be note that use of LSE for a single particle is inappropriate. This is exactly why physicists encountered with these confusing situations use a probabilistic or statistical interpretation of quantum phenomena. On the other hand, LSE can not fully account for dynamical effects of statistical ensemble of particles because it is inherently linear in nature. The NLSE accounting for full potential of a fluid should be used instead of LSE to describe statistical properties of the system. For a quantum fluid consisting of charged particles, ignoring the spin effect, the appropriate NLSE may be written as
\begin{subequations}\label{sp}
\begin{align}
&\left\{ {i\hbar \frac{\partial }{{\partial t}} + \frac{{{\hbar ^2}}}{{2{m}}}\Delta - \left[ {q\Phi  + \mu (\Psi ) + U} \right]} \right\}\Psi  = 0,\\
&\Delta\Phi = 4\pi q\left( {{{\left| \Psi_0  \right|}^2} - {{\left| {{\Psi}} \right|}^2}} \right),
\end{align}
\end{subequations}
in which $\Phi$, $\Psi$ and $\mu(\Psi)$ are the electrostatic potential, the wavefunction and the chemical potential, respectively. Also, $|\Psi|=\sqrt{n}$ with $n$ the local density of particles and equilibrium value $|\Psi_0|^2$ \cite{manfredi}. Note that in the absence of electrostatic interactions $q=0$, (\ref{sp}) does not reduce to the standard LSE due to the nonlinear $\mu(\Psi)$ representing the statistical pressure effects. In the limit of a very dilute classical gas ($\mu\ll 1$) the standard LSE is obtained.

\section{Concluding Remarks}

In this research we investigated the nonlinear excitations in a driven nonlinear Schr\"{o}dinger equation (DNLSE) and showed that pseudoquantum effects observed in recent experiments can be explained in the framework of hydrodynamic fluid model. We also showed that the NLSE fully describes the pseudoquantum nature of the pseudoparticle in the Sagdeev pseudopotential similar to linear Schr\"{o}dinger equation (LSE) which describes the quantum behavior of real particles. We explained many pseudoquantum effects using the DNLSE such as the wave-particle quality, pseudoentanglement, interference and pseudoenergy quantization which are analogous to quantum mechanical features. The hydrodynamic quantum analog
features considered here can explain the recent experiments on the walking droplets accounting for many quantum features ans support the pilot wave theory of quantum mechanics. It was found that, quite similar to pilot wave theory, current hydrodynamic model incorporates the nonlocality which is a characteristic feature of hidden variable theories. We believe that current research on pseudoquantum nature of hydrodynamic excitations can help in better understanding of the foundations of quantum mechanics.

\end{document}